\documentclass{iucr}

     \journalcode{B}

\usepackage{graphicx}
\usepackage{multicol}
\usepackage{multirow}
\usepackage{amsmath}
\usepackage{amssymb}
\usepackage[colorlinks,linkcolor=blue,citecolor=blue]{hyperref}

\begin{document}

\title{Magnetic structure of the noncentrosymmetric magnet Sr$_2$MnSi$_2$O$_7$ through irreducible representation and magnetic space group analyses}
\shorttitle{Magnetic structure of Sr$_2$MnSi$_2$O$_7$}

\cauthor[a,b,c]{Y.}{Nambu}{nambu@tohoku.ac.jp}{}
\author[d]{M.}{Kawamata}
\author[d]{X.}{Pang}
\author[e]{H.}{Murakawa}
\author[f,g]{M.}{Avdeev}
\author[h]{H.}{Kimura}
\author[a]{H.}{Masuda}
\author[e]{N.}{Hanasaki}
\author[a]{Y.}{Onose}

\aff[a]{Institute for Materials Research, Tohoku University, \city{Sendai} 980-8577, \country{Japan}}
\aff[b]{Organization for Advanced Studies, Tohoku University, \city{Sendai} 980-8577, \country{Japan}}
\aff[c]{FOREST, Japan Science and Technology Agency, \city{Saitama} 332-0012, \country{Japan}}
\aff[d]{Department of Physics, Tohoku University, \city{Sendai} 980-8578, \country{Japan}}
\aff[e]{Department of Physics, Osaka University, Toyonaka, \city{Osaka} 560-0043, \country{Japan}}
\aff[f]{Australian Nuclear Science and Technology Organisation, \city{Kirrawee} DC, NSW 2232, \country{Australia}}
\aff[g]{School of Chemistry, University of Sydney, \city{Sydney} 2006, \country{Australia}}
\aff[h]{Institute of Multidisciplinary Research for Advanced Materials, Tohoku University, \city{Sendai} 980-8577, \country{Japan}}

\maketitle

\begin{abstract}
     Magnetic structures of the noncentrosymmetric magnet Sr$_2$MnSi$_2$O$_7$ were examined through neutron diffraction for powder and single-crystalline samples, as well as magnetometry measurements.
     All allowed magnetic structures under the space group $P\bar{4}2_1 m$ with the magnetic wavevector $\vec{q}_{\rm m}=(0,0,1/2)$~r.l.u., were analysed via irreducible representation and magnetic space group approaches.
     The compound is refined to have in-plane magnetic moments within the magnetic space group $Cmc2_1.1^{\prime}_c$ (\#36.177) under zero field, which can be altered to $P2_1 2_1 2_1.1^{\prime}_c$ (\#19.28) above $\mu_0 H=0.067(5)$~T to align induced weak-ferromagnetic components within one layer on the $ab$-plane.
     All refined parameters are provided following the recent framework based upon the magnetic space group, which better conveys when exchanging crystallographic information for commensurate magnetic structures.
\end{abstract}

\section{Introduction}

Multiferroics with coexisting magnetic and ferroelectric orders have garnered significant interest (Dong, 2015).
Magnetically induced ferroelectricity is a promising avenue towards future applications that utilise the mutual control of electricity and magnetism via magnetoelectric coupling.
There have been hitherto proposed three primary mechanisms owing to the spin current (Katsura, 2005), the magnetostriction (Arima, 2006), and the hybridisation between transition-metal $d$ and ligand $p$ orbital (Jia, 2006, Jia, 2007).

The ferroelectricity arising from the spin-dependent $d$-$p$ hybridisation mechanism (Yamauchi, 2011) is typically observable for materials with the point group $\bar{4}2m$ that lacks inversion symmetry.
The magnetically induced ferroelectricity has indeed been demonstrated in CuB$_2$O$_4$ with the space group $I\bar{4}2d$ (Saito, 2008) and Ba$_2$CoGe$_2$O$_7$ with $P\bar{4}2_1 m$ (Murakawa, 2010).
The latter belongs to the melilite group being typical solid-solutions of {\aa}kermanite and gehlenite minerals, which is generally denoted by the chemical formula A$_2$MB$_2$X$_7$ (A = larger alkali earth and lanthanide cations, M, B = smaller divalent to tetravalent cations, X = O, N, F, S anions).

The melilite family provides a wide range of physical phenomena.
In addition to the multiferroic behaviour observed in M = Co$^{2+}$ cases (K\'{e}zsm{\'a}rki, 2011, Endo, 2012, Murakawa, 2012, K\'{e}zsm{\'a}rki, 2014, Kim, 2014, Soda, 2014, Solovyev, 2015), a simple antiferromagnetic transition with Mn$^{2+}$ (Endo, 2012, Sazonov, 2023, Murakawa, 2012) and an incommensurate-commensurate magnetic transition with Cu$^{2+}$ (M{\"u}hlbauer, 2012, Murakawa, 2012, Togawa, 2012, Nucara, 2014, Capitani, 2015) have been detected.
They also attract attention owing to their optical properties (Liang, 2015) and oxygen-conduction (Kuang, 2008).
Besides that, the family is potentially practical for antiferromagnetic spintronics. 
The Dzyaloshinskii-Moriya interaction (DMI) (Dzyaloshinskii, 1958, Moriya, 1960) in noncentrosymmetric systems usually couples the spin degrees of freedom, leading to a rich spin texture in the reciprocal space with applying magnetic fields (Kawano, 2019).
An advantage of having spin texture is that it stores information directly by controlling the texture itself, and it can serve as a prime candidate given the Dresselhaus-type (Dresselhaus, 1955) arrangement of DMI in the family.

In terms of the magnetic moment directions among the melilite family, Ba$_2$MnGe$_2$O$_7$ (Sazonov, 2023), for instance, has in-plane moments along the $ab$-plane, which contrasts with sibling compounds of Ba$_2$MnSi$_2$O$_7$ (Sale, 2019) and Sr$_2$MnGe$_2$O$_7$ (Endo, 2012) having out-of-plane moments.
Given that they have almost identical structures and a difference can only be found in the ionic radii of A$^{2+}$ and B$^{4+}$, microscopic origins for the selected moment direction thus remain unsolved.
To resolve this problem, a further investigation for detailed crystal and magnetic structures for newly developed materials will be required.

In this study, we investigate magnetic structures of the newly developed Sr$_2$MnSi$_2$O$_7$ [structure schematically depicted in Fig.~\ref{fig1}(a)] (Endo, 2010, Koo, 2012, Akaki, 2013).
Powder and single crystalline samples were grown with high quality, and neutron diffraction techniques were applied for both.
Utilising the group theoretical analyses on both powder and single crystal data, magnetic structures under zero and finite fields are successfully refined.
We argue that combining complementary isothermal magnetisation data results in a spin-flip transition accompanying the magnetic space group change.

\section{Experimental}

Polycrystalline samples of Sr$_2$MnSi$_2$O$_7$ were synthesised by the solid-state reaction using the stoichiometric amount of SrCO$_3$, MnO, and SiO$_2$ powders at 1100$^{\circ}$C in the air.
Single crystalline samples were then grown in the air by the floating zone method ($<1$~mm/hour growth speed) using an image furnace (Murakawa, 2012).
The quality of the obtained samples was confirmed by x-ray diffraction measurements.
Isothermal magnetisation was measured using the Physical Property Measurement System manufactured by Quantum Design Inc.

Neutron powder diffraction data were collected on the high-resolution ECHIDNA diffractometer (Avdeev, 2018) at the Australian Nuclear Science and Technology Organisation (ANSTO) with the wavelength, $\lambda=2.4395$~{\AA}.
Diffraction patterns were obtained at $T=1.7$ and 10~K in a cryomagnet without applying a magnetic field.

Single crystal neutron diffraction measurements were performed on the four-circle diffractometer T2-2 FONDER (Noda, 2001), stationed at JRR-3, Japan, with $\lambda=1.2464$~{\AA}.
A magnetic field along the $[H,0,0]$ direction was applied to align magnetic domains using neodymium-based permanent magnets.

\section{Results and discussion}

To refine the magnetic structure of Sr$_2$MnSi$_2$O$_7$, we first performed powder neutron diffraction on ECHIDNA (Avdeev, 2018).
Figure~\ref{fig2}(a) depicts the raw data taken at $T=10$ and 1.7~K, where additional resolution-limited peaks associated with magnetic ordering are visible at 1.7~K below the N\'eel temperature, $T_{\rm N}=3.4$~K (Endo, 2010, Akaki, 2013).
These diffraction patterns are well fit based on Sr$_2$MnSi$_2$O$_7$ and aluminium coming from the cryomagnet, which are treated as Le Bail phases.

\begin{figure}
     \caption{(a) Crystallographic unit cell of Sr$_2$MnSi$_2$O$_7$ with arrows marking each element. Refined magnetic structure of Sr$_2$MnSi$_2$O$_7$ under (b) $\mu_0 H=0$~T using powder and (c) $\sim 0.31$~T using single crystalline samples. The magnetic unit cell is $a\times b \times 2c$ reflecting the magnetic wavevector, $\vec{q}_{\rm m}=(0,0,1/2)$~r.l.u.}
     \includegraphics[width=0.8\linewidth,bb=0 0 485 538]{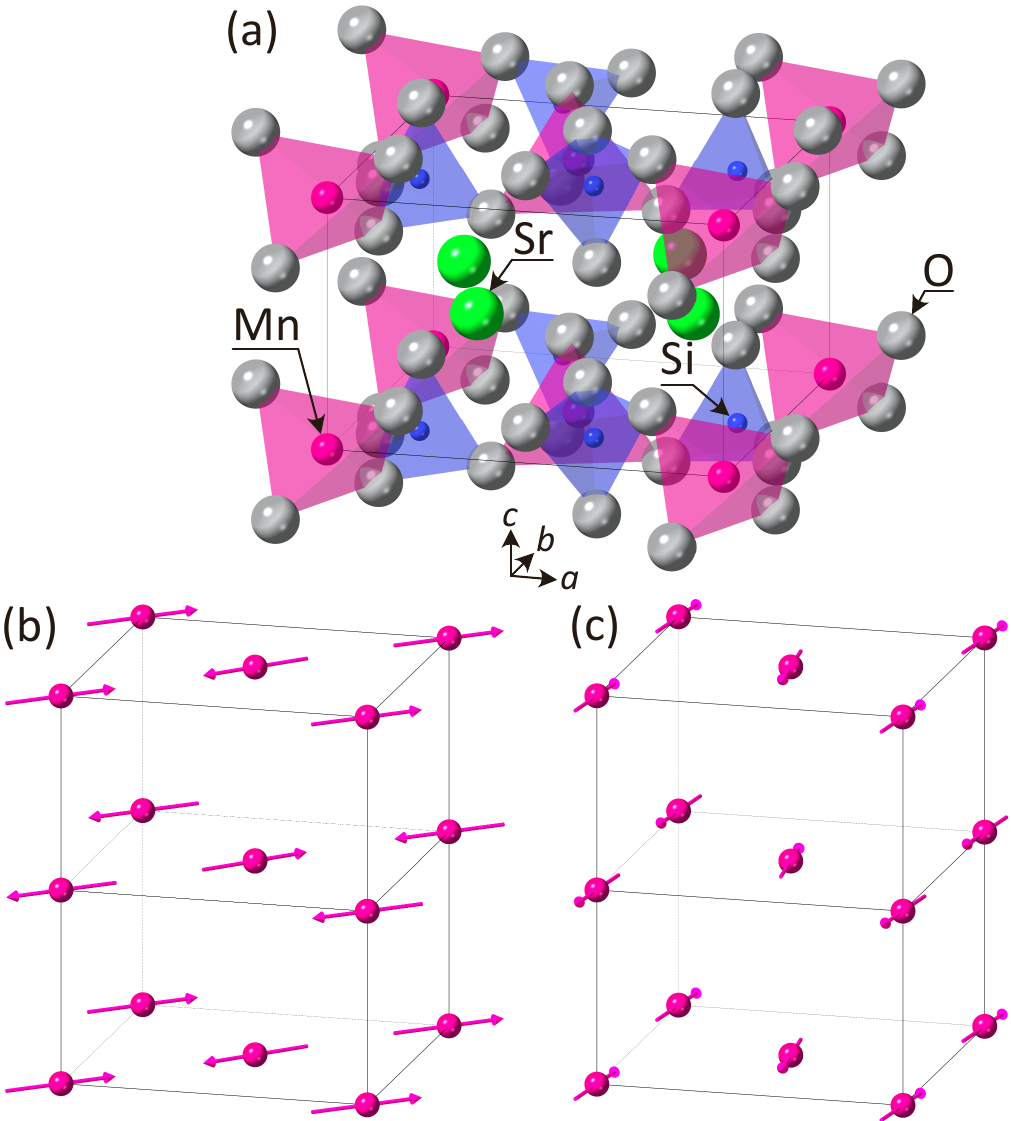}
     \label{fig1}
\end{figure}

\begin{figure}
     \caption{Neutron powder diffraction patterns of Sr$_2$MnSi$_2$O$_7$ collected on the powder diffractometer ECHIDNA. (a) Raw data taken at $T=10$ and 1.7~K with shaded areas denoting purely magnetic contributions, and data for (b) 10~K and (c) 1.7~K with Rietveld refinement (solid lines). The calculated positions of nuclear and magnetic reflections, and aluminium reflections from the cryomagnet treated as Le Bail phases, are indicated (green ticks). The bottom lines give the difference between observed and calculated intensities.}
     \includegraphics[width=\linewidth,bb=0 0 897 913]{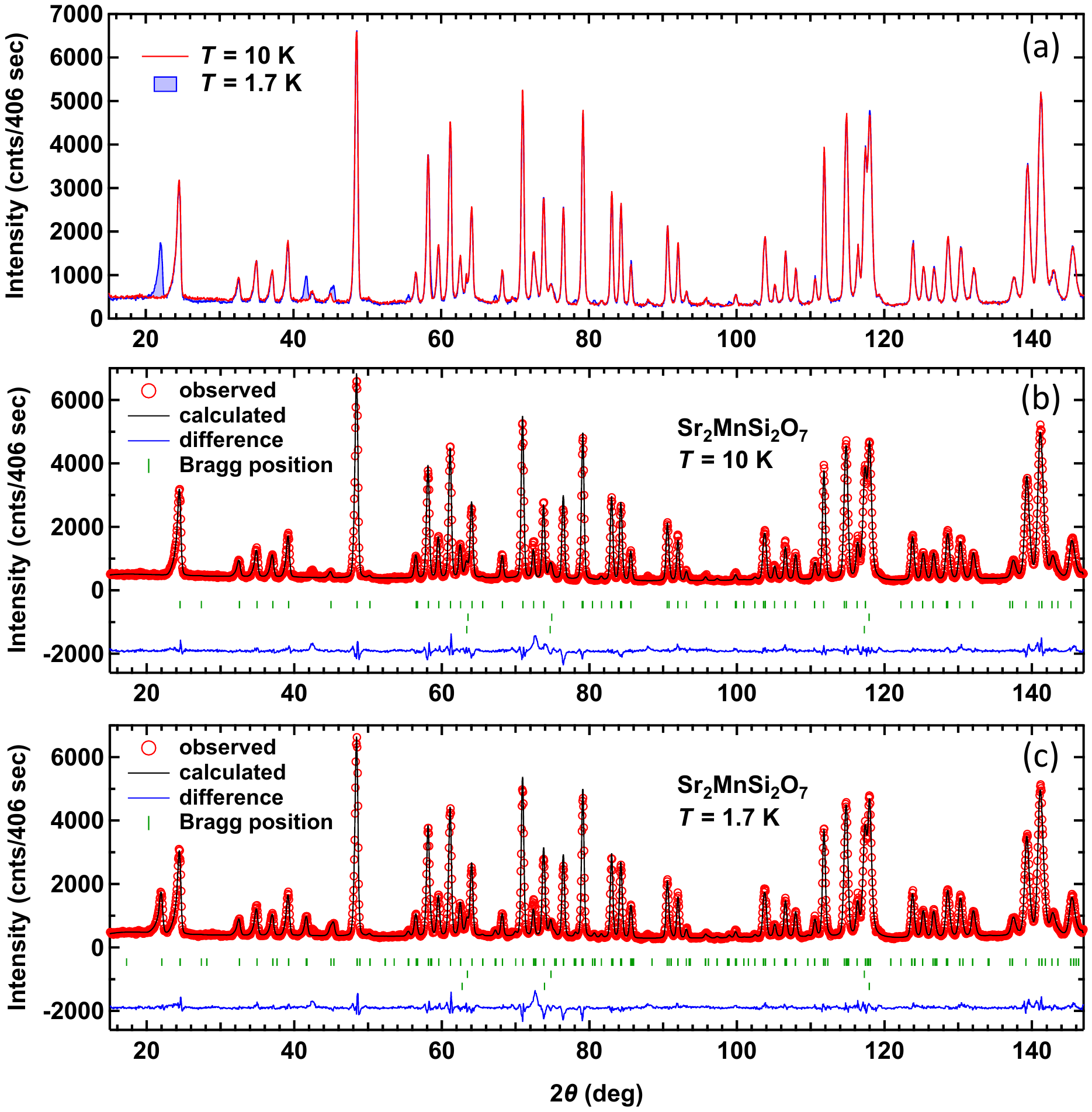}
     \label{fig2}
\end{figure}

Figure~\ref{fig1}(a) schematically illustrates the crystal structure of Sr$_2$MnSi$_2$O$_7$.
The structure comprises corner-shared MnO$_4$ and SiO$_4$ tetrahedra forming layers on the $ab$-plane, separated by Sr$^{2+}$ cations.
Magnetism is associated with a high-spin state of Mn$^{2+}$ ($3d^5$) spins, $S=5/2$, which makes square-planar arrangements.
Table~\ref{table1} summarises the structural parameters of the paramagnetic phase at $T=10$~K determined by the Rietveld refinement [Fig.~\ref{fig2}(b)].
For the 2.4395~{\AA} wavelength, the accessible momentum-transfer ($Q$) regime is quite limited ($Q\le 4.9$~{\AA}$^{-1}$), which was insufficient to refine the isotropic Debye-Waller factor ($B_{\rm iso}$) for this structure.
During the analyses, the $B_{\rm iso}$ value was thus fixed to be 0.1~{\AA}$^2$.
Consistent with the earlier reports (Endo, 2010, Akaki, 2013), the crystal structure [Fig.~\ref{fig1}(a)] is accounted for by the tetragonal space group $P\bar{4}2_1 m$.
By keeping the site occupancy variable for each atom, the result indicates the approximately stoichiometric ratio of Sr$_{1.99(1)}$Mn$_{1.03(1)}$Si$_{2.05(1)}$O$_{7.02(3)}$.

We here apply representation analysis to identify the magnetic structure.
All the magnetic peak positions can be accounted for by the single magnetic wavevector, $\vec{q}_{\rm m}=(0,0,1/2)$~r.l.u. [Fig.\ref{fig2}(c)], which denotes magnetic unit cell doubled along the $c$-axis.
Basis vectors (BVs) of the irreducible representations (irreps) for the wavevector with the Kovalev notation (Kovalev, 1965, Kovalev, 1993) are summarised in Table~\ref{table2} together with schematic drawings of magnetic structures corresponding to each BV.
There are six BVs in total, belonging to three distinct irreps: one-dimensional $\Gamma_1$, $\Gamma_2$, and two-dimensional $\Gamma_5$.
The $\psi_1$ within $\Gamma_1$ and $\psi_2$ in $\Gamma_2$ stand for the collinear G-type antiferromagnetic and ferromagnetic spin arrangements, respectively, where spins are aligned along the $c$-axis.
The $\Gamma_5$ containing four BVs, on the other hand, allows spins projected onto the $ab$-plane.

To refine the magnetic structure, every possible BV allowed by symmetry is tested by comparing $R_{\rm mag}$ factors.
The BVs within $\Gamma_1$ and $\Gamma_2$ poorly reproduce the obtained data ($R_{\rm mag}>20$\%).
In the case of the two-dimensional $\Gamma_5$, two BVs have to join to form magnetic structures.
The best fit appears to be $R_{\rm mag}=8.76$\% with the combination of $\psi_3$ and $\psi_4$.
The second comes with 9.33\% for $\psi_5$ and $\psi_6$, whereas all the remaining possible combinations exceed 40\%.
We anticipate that the transition at $T_{\rm N}$ is of second order, thus, Landau's theory (Landau, 1984) holds that only one irrep is involved. 
The 1.7~K ($=0.5T_{\rm N}$) magnetic structure corresponding to the best fit is illustrated in Fig.~\ref{fig1}(b), and the estimated moment is $3.17(34)$~$\mu_{\rm B}$ per an Mn$^{2+}$ site.
The moment size is remarkably smaller than Ba$_2$MnSi$_2$O$_7$ ($4.1(1)~\mu_{\rm B}$ at $0.53T_{\rm N}$) (Sale, 2019) and Sr$_2$MnGe$_2$O$_7$ ($3.99(5)~\mu_{\rm B}$ at $0.57T_{\rm N}$) (Endo, 2012) with out-of-plane moments, yet is comparable with Ba$_2$MnGe$_2$O$_7$ ($3.24(3)~\mu_{\rm B}$ at $0.63T_{\rm N}$) (Sazonov, 2023) with in-plane moments.
The refined magnetic structure clearly breaks $\bar{4}$ symmetry of the parent space group, implying that the actual space group below $T_{\rm N}$ is no longer tetragonal.
Reflecting the refined coefficients of $\psi_3$ and $\psi_4$, which turn out to be similar in magnitude, spins are primarily pointing to the $[110]$ and $[\bar{1}\bar{1}0]$ directions [Fig.~\ref{fig1}(b)] yet the slight deviation yields a marginally canted arrangement.
The canting leads to weak ferromagnetic components along the $[1\bar{1}0]$ direction, which are restricted within one layer on the $ab$-plane but stacks antiferromagnetically along the $c$-axis, yielding no net magnetisation, as will be discussed later on.
Note that the moments are in-plane but the direction within the plane cannot be precisely determined from powder data given no significant orthorhombic distortion is detected.

We also adopt the analysis based upon the magnetic space group (MSG).
Allowed maximal MSGs for the space group $P\bar{4}2_1 m$ with the wavevector $(0,0,1/2)$~r.l.u. are six in total, in which two of them particularly exclude moments at the Mn sites.
The rest of the four with the Uniﬁed (UNI) symbols (Campbell, 2022) is schematically depicted in Table.~\ref{table3}.
The MSGs $P\bar{4}2_1 c.1^{\prime}_c$ (\#114.280) and $P\bar{4}2_1 m.1^{\prime}_c$ (\#113.272) correspond to antiferromagnetic structures with spins along the $c$-axis, where spin arrangements on the $ab$-plane are ferromagnetic and antiferromagnetic, respectively.
They are identical to $\psi_2$ in $\Gamma_2$ and $\psi_1$ in $\Gamma_1$ from the representation analysis.
With the in-plane moments cases, $Cmc2_1.1^{\prime}_c$ (\#36.177) perfectly matches with the combination of $\psi_3$ and $\psi_4$ within $\Gamma_5$, and $P2_1 2_1 2_1.1^{\prime}_c$ (\#19.28), when $m_x=m_y$ in particular, corresponds to $\psi_3\ominus\psi_5$, where the coefficients need to be restricted to the same magnitude.

We sort out all the four possibilities to refine the magnetic structure by again comparing $R_{\rm mag}$ factors.
Consistent with the results from the above-mentioned representation analysis, refinement based upon $Cmc2_1.1^{\prime}_c$ gives the best fit ($R_{\rm mag}=8.78$\%), being the same MSG in Ba$_2$FeSi$_2$O$_7$ (Jang, 2021).
$P2_1 2_1 2_1.1^{\prime}_c$ comes second with 9.66\%, the others cannot reproduce the data ($R_{\rm mag}>20$\%).
Thus, both irrep and MSG approaches consistently give the same refined magnetic structure, with moment sizes that are consistent with each other.

Magnetic moments mostly parallel to $[H,H,0]$ are arranged to form the G-type antiferromagnetic structure, and the moments for each Mn layer stack antiferromagnetically [Fig.~\ref{fig1}(b)].
The induced weak ferromagnetic component per layer is allowed to have four sorts of domains reflecting the rotational symmetry inherent, and the resultant tilting angle from the $[H,H,0]$ direction [defined as $\phi$ in Fig.~\ref{fig4}] may have been underestimated owing to the spherically $\vec{Q}$-averaged characteristic of the powder diffraction.
The canting can be induced by the out-of-plane component of DMI, $D_z$, arising from the noncentrosymmetric structure.
Through the variational method on the spin Hamiltonian, including the nearest-neighbour interaction, $J_1$, and $D_z$, the following relation can be derived,
\begin{align}
     \frac{D_z}{J_1}=\tan 2\phi.
\end{align}
The refined $\phi$ angle of $\sim 2.4$~deg poses the lower minimum of the relation, $D_z/J_1=0.09$.

To minimise the effects of the magnetic domain formation, we have conducted single crystal neutron diffraction on FONDER (Noda, 2001) by applying the magnetic field of $\mu_0 H=0.31$~T on average at the sample position, where the applied fields were evaluated using a Gaussmeter, giving 0.21~T and 0.40~T at the top and bottom of the sample positions, respectively.
The field was set along the $a$-axis, which was also selected as the longitudinal direction of the Gifford-McMahon-type cryostat, and the single crystalline sample was aligned on the nominal $[0,K,L]$ scattering zone.
An initial guess can be assigned that the weak ferromagnetic component on a layer naturally rotates and becomes parallel to $\vec{H}\parallel a$, yielding spin arrangements mostly along the $b$-axis.

\begin{figure}
     \caption{The relation between observed and calculated structure factors for (a) crystal [$R_F=3.70$\%] and (b) magnetic structure [$R_F=10.65$\%] collected on the four-circle diffractometer FONDER taken at 2.8(3)~K. The inset to (a) is a photo of the single crystalline sample and its neighbours.}
     \includegraphics[width=\linewidth,bb=0 0 686 327]{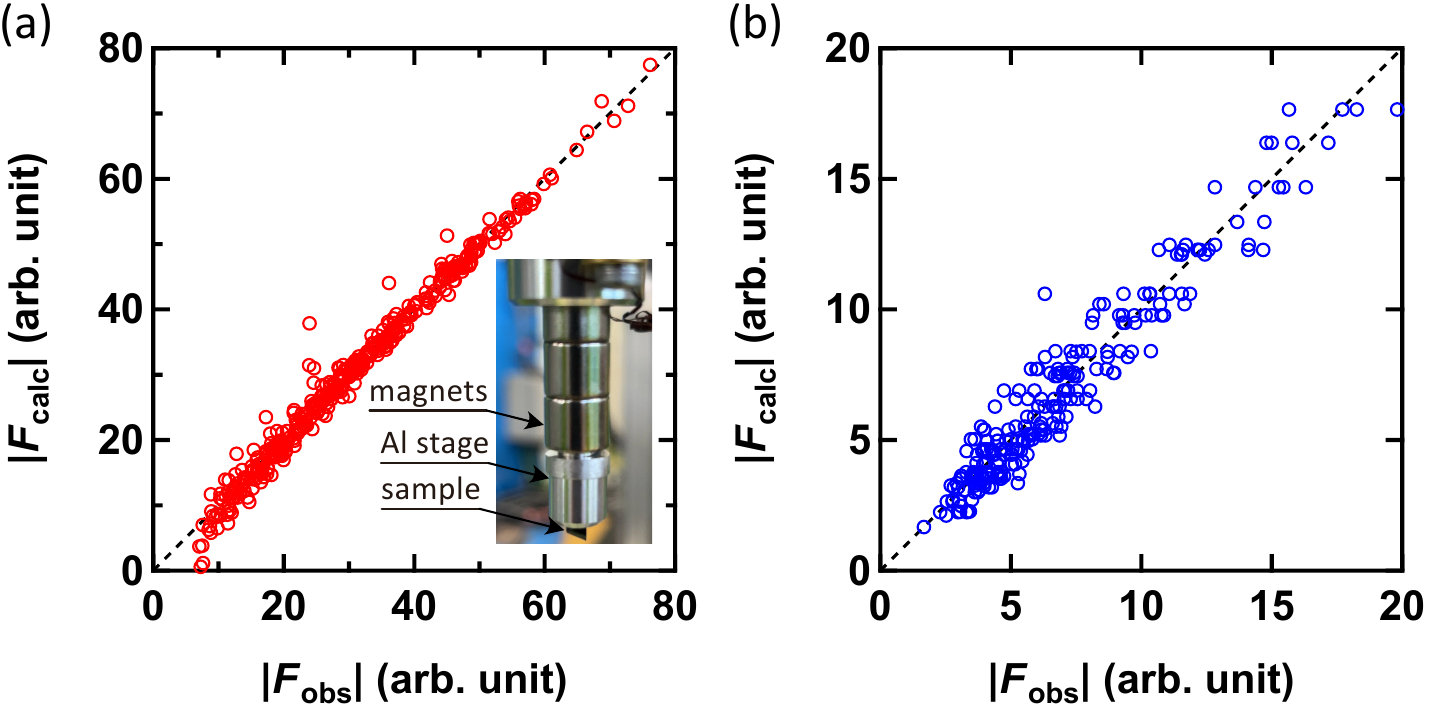}
     \label{fig3}
\end{figure}

\begin{figure}
     \caption{(a) Isothermal magnetisation of Sr$_2$MnSi$_2$O$_7$ taken at 2~K with $\mu_0 H\parallel a$ up to $\mu_0 H=9$~T. (b) Isothermal magnetisation (left-axis) for lower fields. The first derivative (right-axis) shows a small hump at $\mu_0 H=0.067(5)$~T, where the $\pi/4$ spin flip accompanying the magnetic space group change from $Cmc2_1.1^{\prime}_c$ to $P2_1 2_1 2_1.1^{\prime}_c$ takes place. The closed and open arrows depict the directions of the magnetic field and induced weak-ferromagnetic components per layer with schematic spin configurations, respectively.}
     \includegraphics[width=\linewidth,bb=0 0 1316 503]{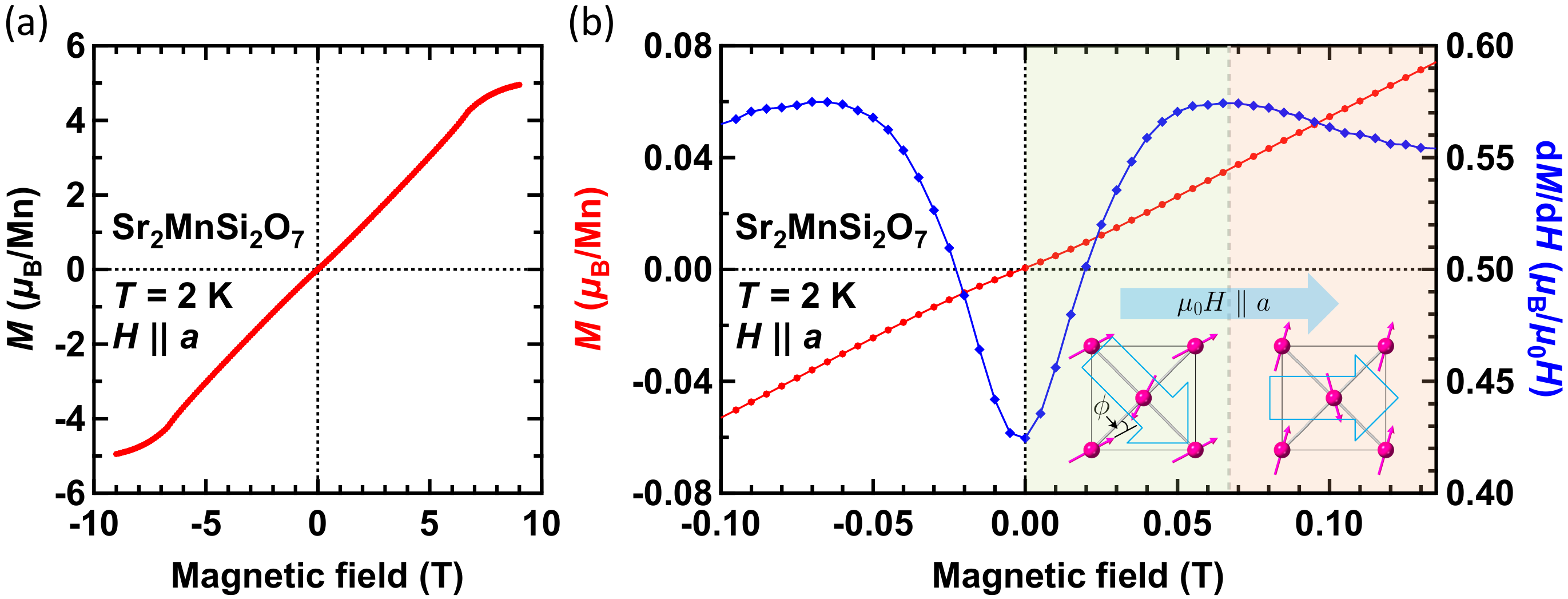}
     \label{fig4}
\end{figure}

Measurements for collecting nuclear and magnetic reflections were performed at the lowest temperature of the cryostat.
The temperature stability turned out to be insufficient, reflecting a large heat capacity of the permanent magnets themselves and added covering materials [inset to Fig.~\ref{fig3}(a)].
The temperature warmed typically when the longitudinal direction was tilted (greater $\chi$-angle) towards the reflections with the finite $H$-index. 
The fact led to the relatively considerable uncertainty of the measured temperature, 2.8(3)~K, making the magnetic moment sizes also fluctuate with approaching $T_{\rm N}$.
We therefore selected only data meeting the $|\chi|\le 40$~deg criteria for magnetic structure refinement.
Obtained data were corrected for the Lorentz factor and the absorption depending upon the neutron flight path lengths within the sample, and then converted into the structure factor, $|F_{\rm obs}|$. 

For the crystal structure refinement, 429 reflections were collected in total to estimate the scale factor and the linear coefficient of the isotropic extinction correction.
Refinements give a nicely fit proportional relation between the observed and calculated structure factors, as depicted in Fig.~\ref{fig3}(a).
We then measured as many magnetic reflections as possible and only used $|\chi|\le 40$~deg reflections for analysis to avoid temperature increases, which counted up to 265 reflections.
We compare $R_F$ factors for all the possible MSGs, giving $P2_1 2_1 2_1.1^{\prime}_c$ with $R_F=10.65$\% [Fig.~\ref{fig3}(b)] against $Cmc2_1.1^{\prime}_c$ with 15.04\%.
Figure~\ref{fig1}(c) illustrates the refined magnetic structure within $P2_1 2_1 2_1.1^{\prime}_c$, where the moments are mainly pointing to the $b$-axis with a slight canting along the $[H,0,0]$ direction as anticipated.

Interestingly, the magnetic structure has altered from zero field by applying such a small field.
To understand this behaviour, isothermal magnetisation at 2~K was measured.
Figure~\ref{fig4}(a) shows the magnetisation curve under the field along the $[H,0,0]$ direction, and its first derivative (${\rm d}M/{\rm d}H$) under low fields [Fig.~\ref{fig4}(b)] shows a small hump at $\mu_0 H=0.067(5)$~T.
This is reminiscent of the Sr$_2$CoSi$_2$O$_7$ case (Nishina, 2015), whose ${\rm d}M/{\rm d}H$ has a similar hump at one order higher field, 0.7~T.
Given that Sr$_2$CoSi$_2$O$_7$ is considered to demonstrate $\pi/4$ spin-flip from the ground state with mostly $[H,0,0]$ directional moments to $[H,H,0]$ (Nishina, 2015), our compound also possibly has a similar flipping.
Under the finite field, the tilting angle deviated from the $[0,K,0]$ direction is estimated to be $\sim 5.7$~deg being enhanced compared to the powder data under zero field, indicating that aligning magnetic domains by the field indeed works and the tilting is further enhanced via the Zeeman effect.
To quantitatively estimate the strength of $D_z$ under fields, the absolute values of $J_1$ and further neighbour interactions are required, which will be reported elsewhere.

Finally, the refined magnetic structures for zero ($Cmc2_1.1^{\prime}_c$) and finite ($P2_1 2_1 2_1.1^{\prime}_c$) magnetic fields are summarised in Table~\ref{table4}.
Note that $Cmc2_1.1^{\prime}_c$ is allowed to possess an out-of-plane electric polarisation, whereas in $P2_1 2_1 2_1.1^{\prime}_c$, such an effect is forbidden by symmetry.
The crystallographic information under MSGs is a straightforward standalone description.
In contrast, the representation in terms of spin basis vectors stands for a comparison against a parent structure in the paramagnetic phase.
Parameters in Table~\ref{table4} are described within the MSG following the recently suggested guidelines by the International Union of Crystallography (IUCr) Commission on Magnetic Structures, together with the needed information, including the positions of nonmagnetic atoms.
The low-temperature space group under zero field is plausibly $Cmm2$ that makes it possible to have separate lattice constants $a$ and $b$.
However, this orthorhombicity, which is actually fittable, appears quite small and subtle, falling within the experimental accuracy.
Therefore, further experimental probes will be required to fully refine the true space group.

\section{Conclusions}

To summarise, we have refined the magnetic structures of the noncentrosymmetric antiferromagnet Sr$_2$MnSi$_2$O$_7$.
Using powder under zero field and single crystalline samples with a finite field, the magnetic structures are successfully refined based upon parallel approaches using irreducible representation and magnetic space group analysis.
Combining the group theoretical analyses and isothermal magnetisation data, the zero field structure within $Cmc2_1.1^{\prime}_c$ can be switched to $P2_1 2_1 2_1.1^{\prime}_c$ at $\mu_0 H=0.067(5)$~T.
The refined crystalographic information is provided under the magnetic space group framework following the IUCr Commission on Magnetic Structures, serving as a fine example for communicating commensurate magnetic structures. 

\ack{Acknowledgements}

We thank T. Arima, H. Kawamura, T. Koretsune, and Y. Noda for their valuable discussions.
This work was supported by the JSPS (Nos.~21H03732, 22H05145, 17H06137, 24K00572, 24H01638, 24H00189), FOREST (No.~JPMJFR202V) and SPRING from JST, and the Graduate Program in Spintronics and the Graduate Program in Materials Science at Tohoku University.

\clearpage

\begin{table}
     \caption{Atomic positions of Sr$_2$MnSi$_2$O$_7$ at $T=10.0(2)$~K within the space group $P\bar{4}2_1 m$ determined by Rietveld analysis ($R_F=3.34$\%, $\chi^2=4.43$). Lattice constants are $a=b=8.1215(2)$~{\AA} and $c=5.1487(1)$~{\AA}. Fixed isotropic Debye-Waller factor, $B_{\rm iso}=0.1$~{\AA}$^2$ (see main text), is employed.}
     \begin{tabular}{lcccc}
     Atom & Site    & $x$          & $y$          & $z$          \\
     \hline
     Sr1  & $4e$    & 0.3350(2)    & 0.1650(2)    & 0.5045(5)    \\
     Mn1  & $2a$    & 0            & 0            & 0            \\
     Si1  & $4e$    & 0.1369(3)    & 0.3631(3)    & 0.9351(7)    \\
     O1   & $2c$    & 1/2          & 0            & 0.1663(8)    \\
     O2   & $4e$    & 0.1381(2)    & 0.3619(2)    & 0.2523(6)    \\
     O3   & $8f$    & 0.0793(2)    & 0.1939(2)    & 0.7924(4)    \\
     \end{tabular}
     \label{table1}
\end{table}

\begin{table}
     \caption{Basis vectors (BVs) of irreducible representations (irreps) for the space group $P\bar{4}2_1 m$ with the magnetic wavevector $\vec{q}_{\rm m}=(0,0,1/2)$~r.l.u. The atoms are defined as \#1: $(0,0,0)$ and \#2: $(1/2,1/2,0)$.}
     \begin{tabular}{ccccccccl}
                    &              & \multicolumn{3}{c}{atom~\#1}   & \multicolumn{3}{c}{atom~\#2}  &   \\
     irrep          & BV           & $m_x$   & $m_y$   & $m_z$   & $m_x$   & $m_y$   & $m_z$   &   \\
     \hline
     $\Gamma_1$     & $\psi_1$     & 0       & 0       & 4       & 0       & 0       & -4 &
     \begin{minipage}{50mm}
          \scalebox{0.085}{\includegraphics[bb=0 0 842 617]{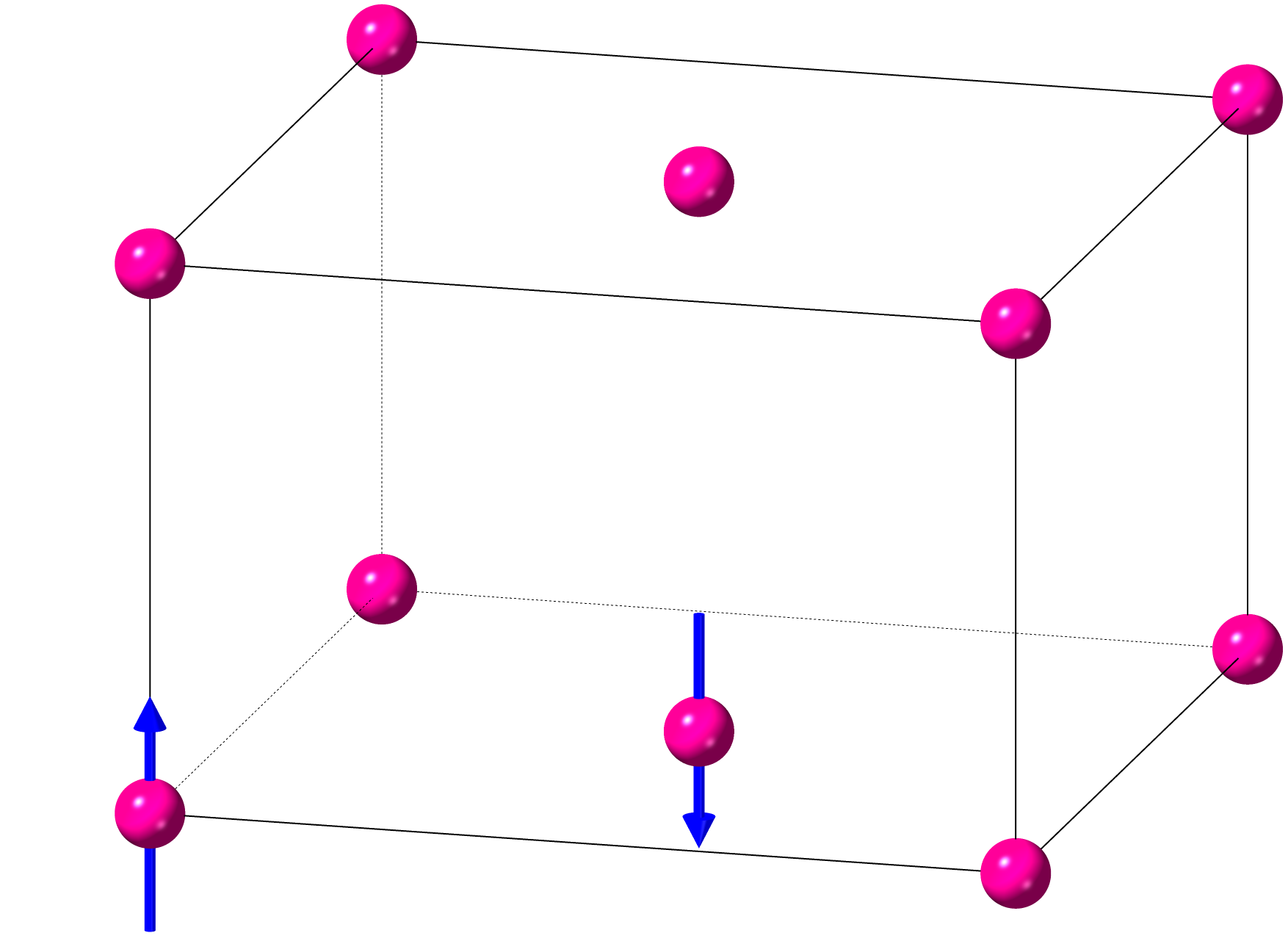}}
     \end{minipage}
     \\
     $\Gamma_2$     & $\psi_2$     & 0       & 0       & 4       & 0       & 0       & 4  &
     \begin{minipage}{50mm}
          \scalebox{0.085}{\includegraphics[bb=0 0 842 617]{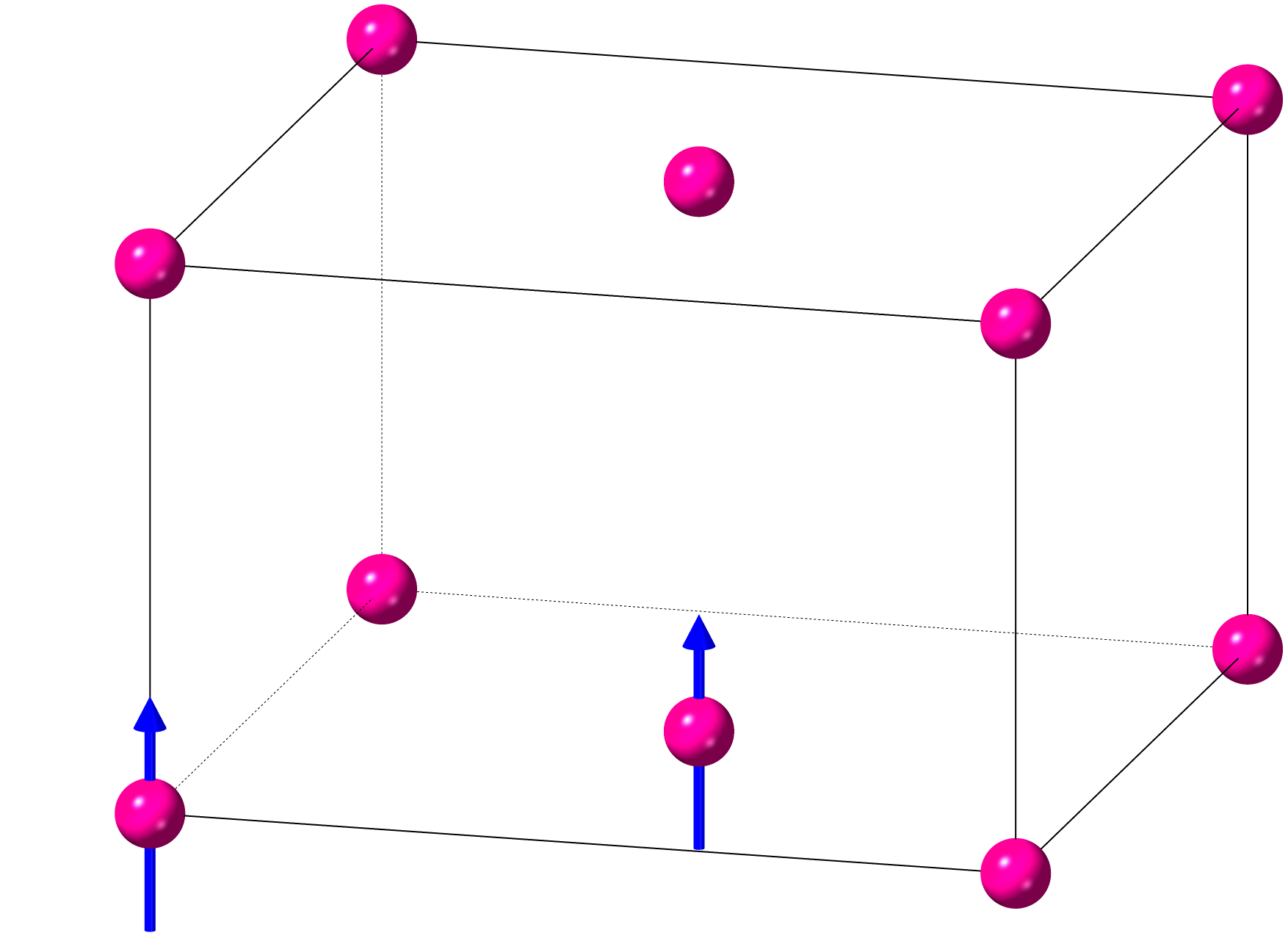}}
     \end{minipage}
     \\
     $\Gamma_5$     & $\psi_3$     & 2       & 0       & 0       & 0       & -2      & 0     &
     \begin{minipage}{50mm}
          \scalebox{0.085}{\includegraphics[bb=0 0 842 617]{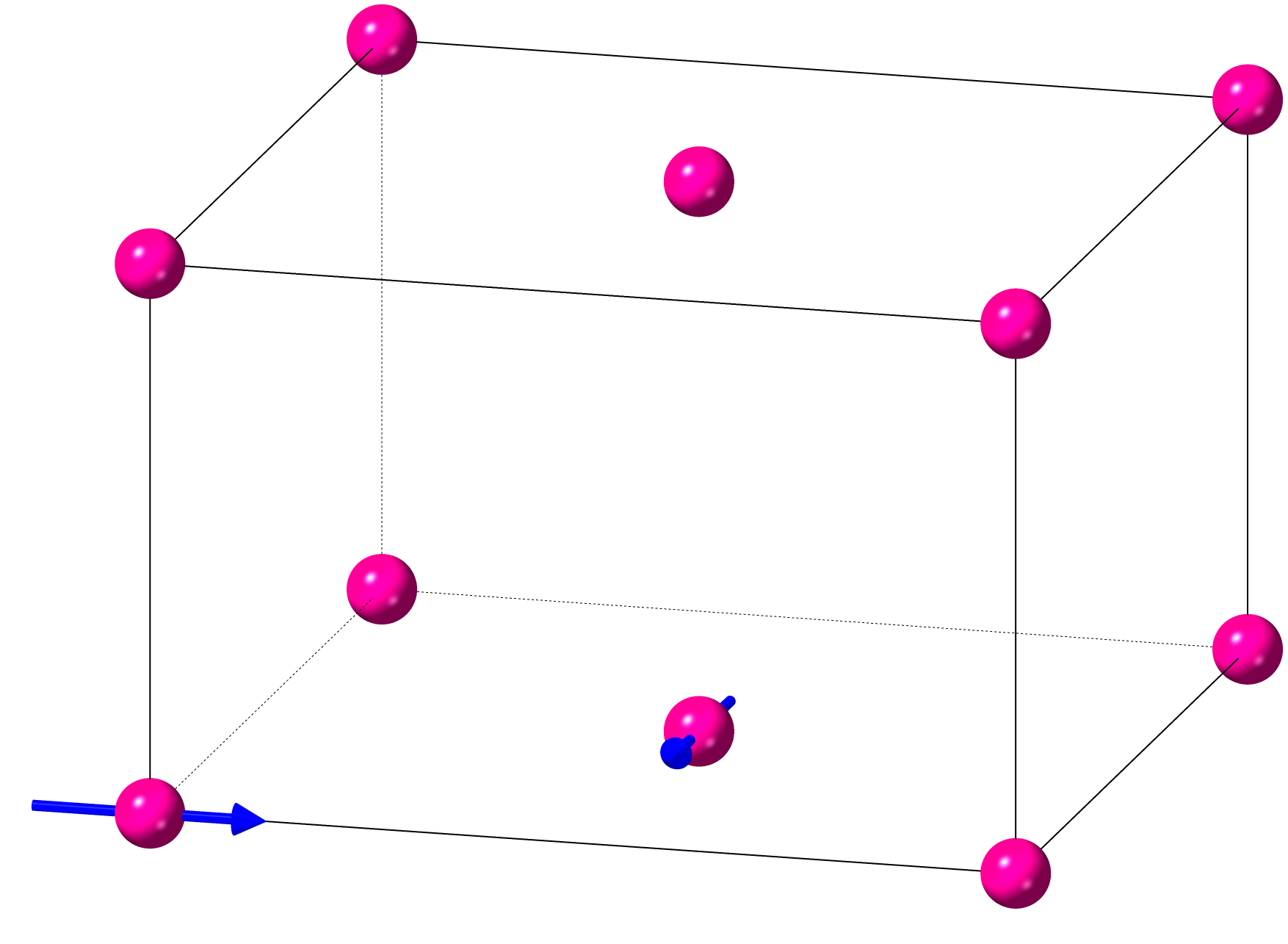}}
     \end{minipage}
     \\
     $\Gamma_5$     & $\psi_4$     & 0       & 2       & 0       & -2      & 0       & 0  &
     \begin{minipage}{50mm}
          \scalebox{0.085}{\includegraphics[bb=0 0 842 617]{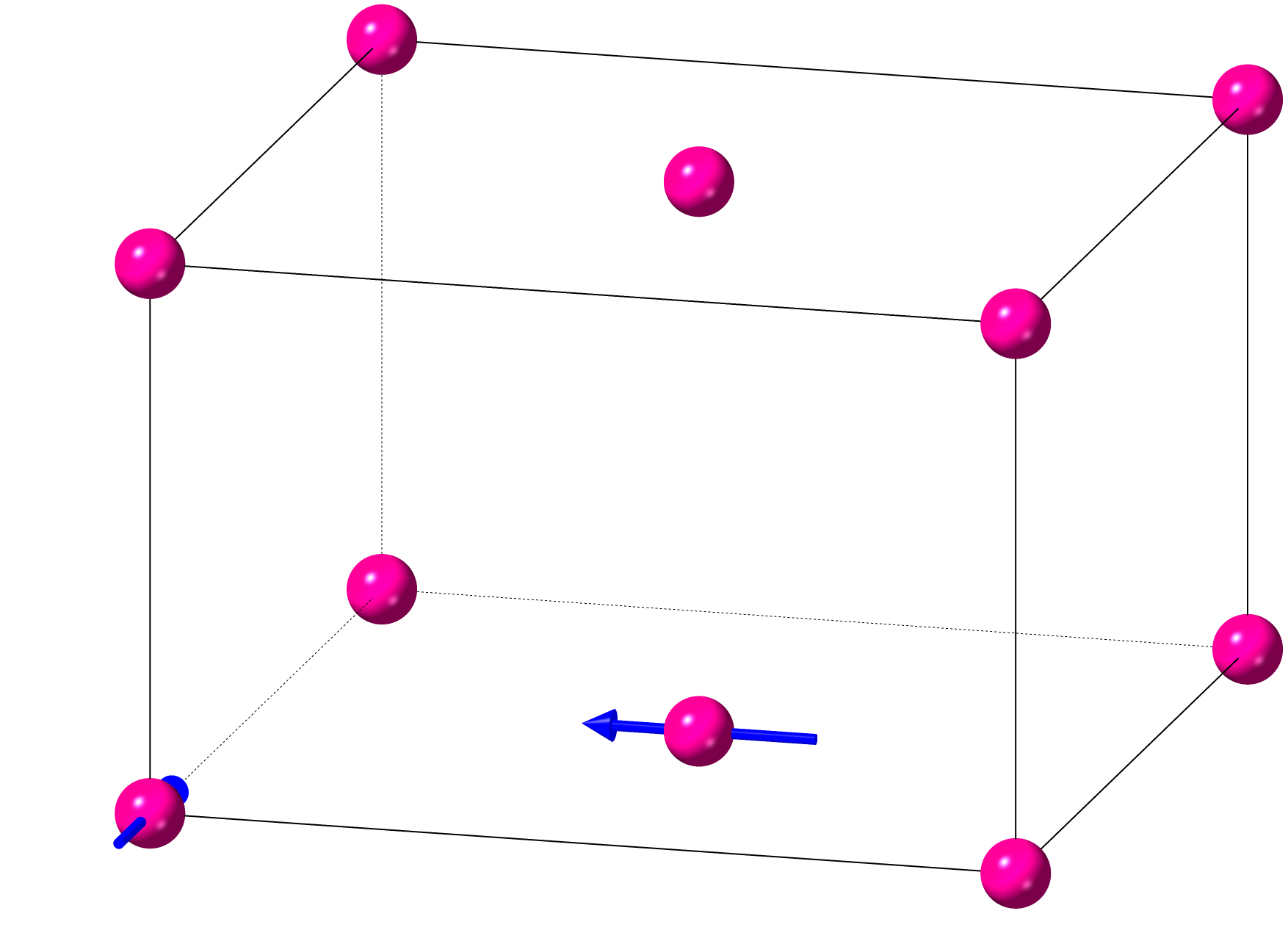}}
     \end{minipage}
     \\
     $\Gamma_5$     & $\psi_5$     & 0       & -2      & 0       & -2      & 0       & 0  &
     \begin{minipage}{50mm}
          \scalebox{0.085}{\includegraphics[bb=0 0 842 617]{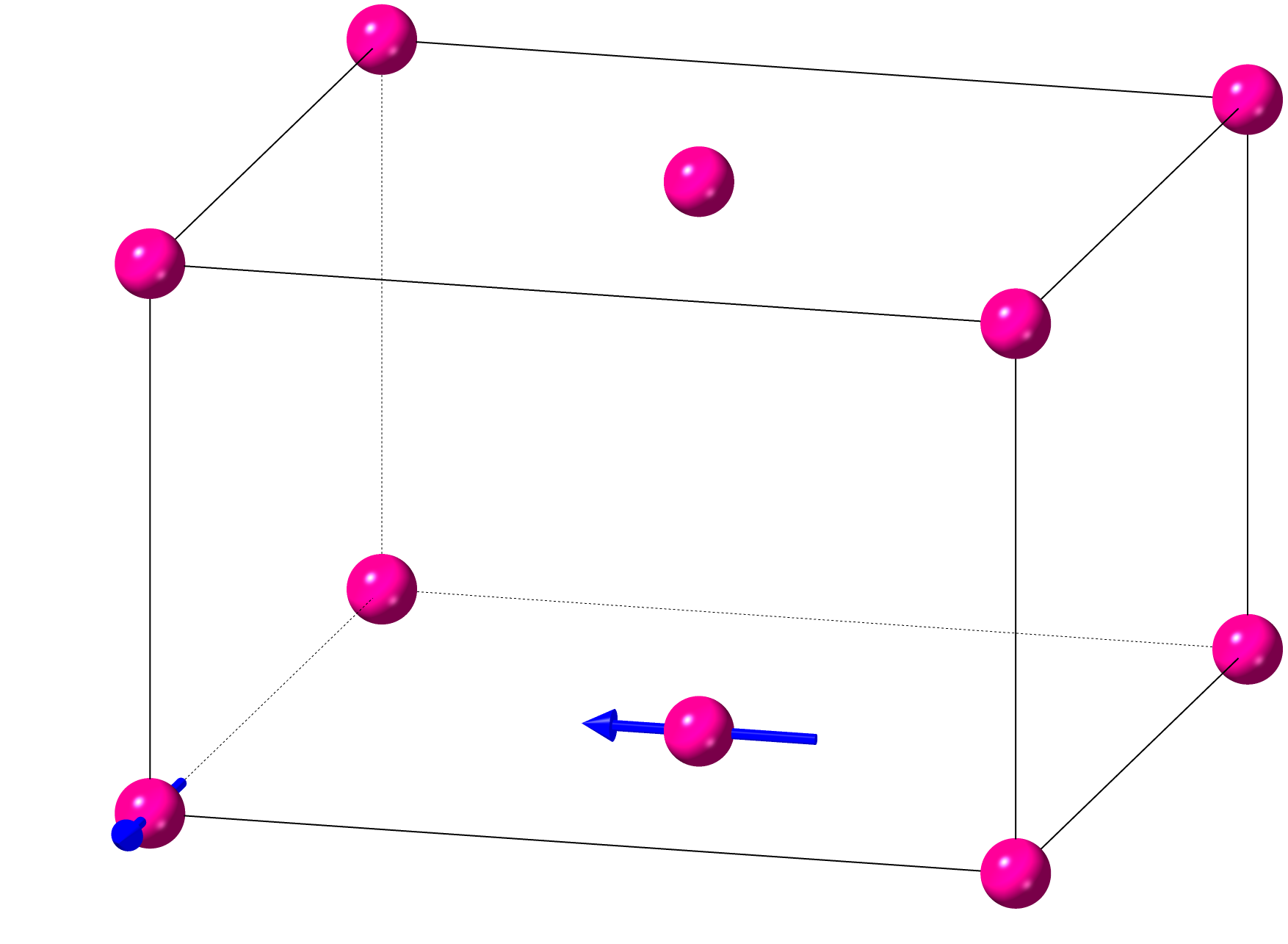}}
     \end{minipage}
     \\
     $\Gamma_5$     & $\psi_6$     & 2       & 0       & 0       & 0       & 2       & 0  &
     \begin{minipage}{50mm}
          \scalebox{0.085}{\includegraphics[bb=0 0 842 617]{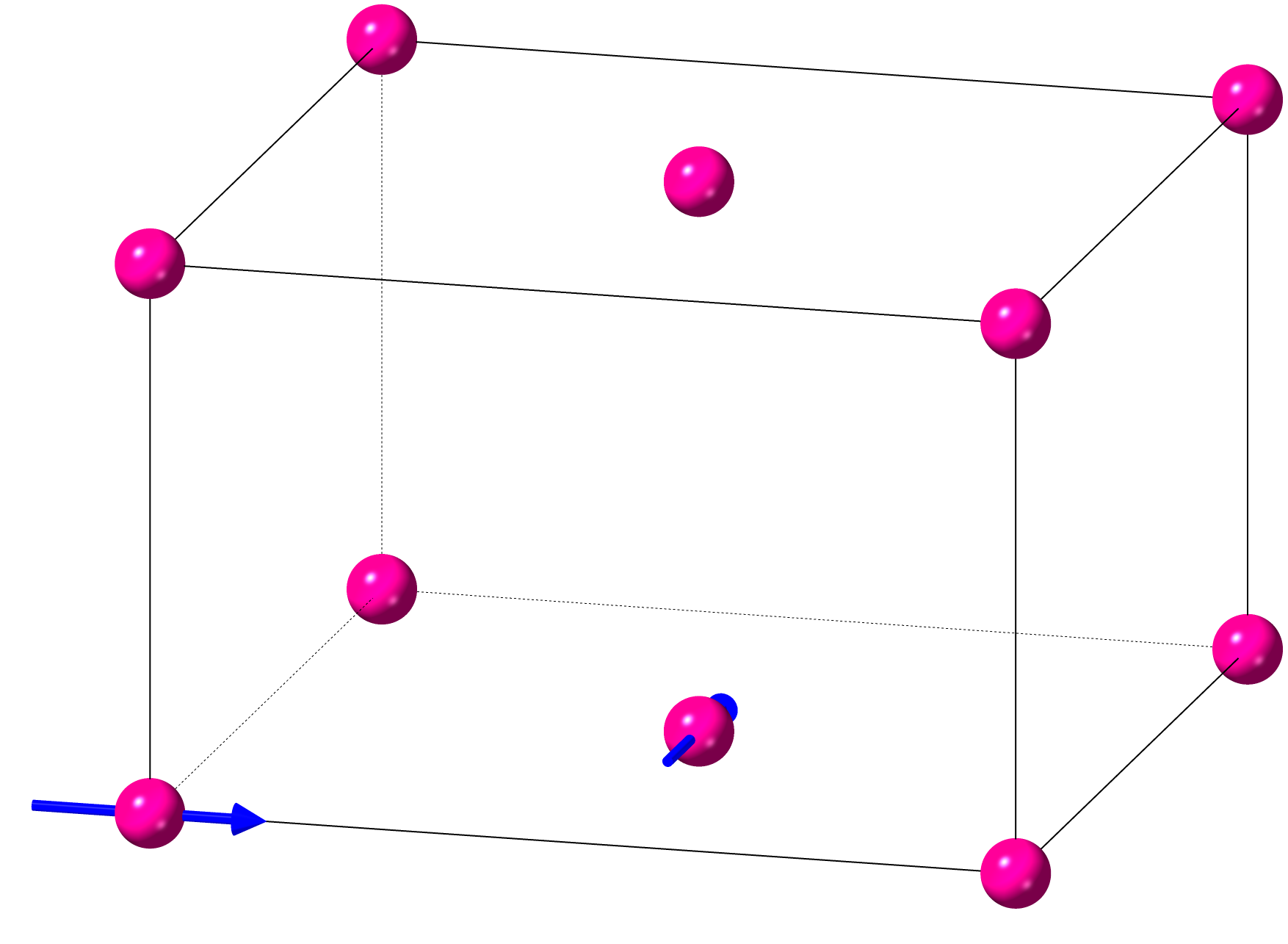}}
     \end{minipage}
     \\
     \end{tabular}
     \label{table2}
\end{table}

\onecolumn

\begin{table}
     \caption{General positions of the maximal magnetic space groups (MSGs) of the parent space group $P\bar{4}2_1 m$ with the magnetic wavevector $\vec{q}_{\rm m}=(0,0,1/2)$~r.l.u. The Uniﬁed (UNI) symbols (Campbell, 2022) in the basis of $(\vec{a},\vec{b},2\vec{c};0,0,0)$ is employed. Each MSG has its corresponding number as $P\bar{4}2_1 c.1^{\prime}_c$ (\#114.280), $P\bar{4}2_1 m.1^{\prime}_c$ (\#113.272), $Cmc2_1.1^{\prime}_c$ (\#36.177), and $P2_1 2_1 2_1.1^{\prime}_c$ (\#19.28).}
     \begin{tabular}{cccl}
     MSG  & Mn positions & magnetic moment   &   \\
     \hline
     $P\bar{4}2_1 c.1^{\prime}_c$   & $(0,0,0 | 0,0,m_z)$ $(1/2,1/2,0 | 0,0,m_z)$ $(0,0,1/2 | 0,0,-m_z)$ $(1/2,1/2,1/2 | 0,0,-m_z)$   & $(0,0,M_z)$  &     
     \begin{minipage}{50mm}
          \scalebox{0.085}{\includegraphics[bb=0 0 842 617]{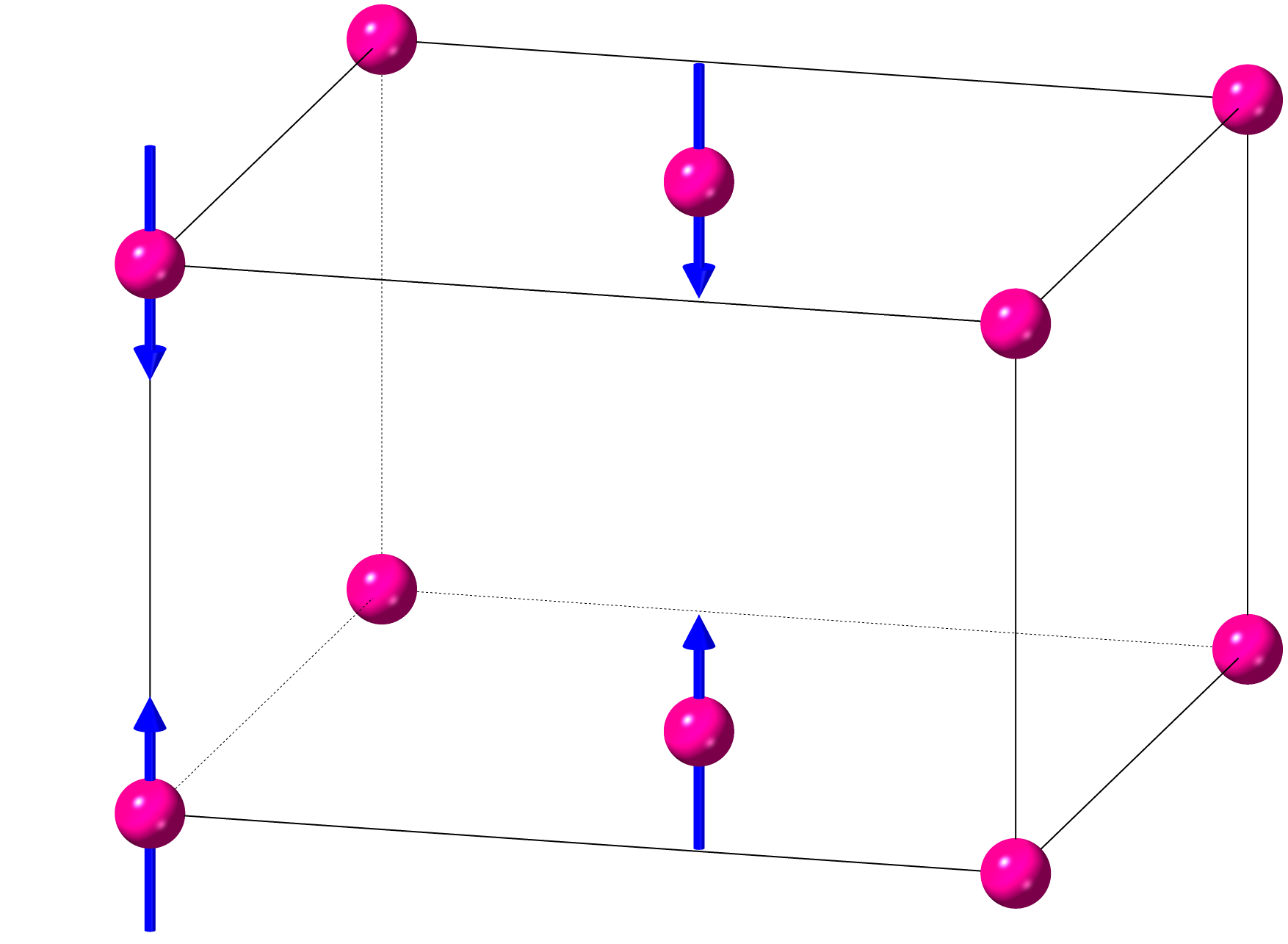}}
     \end{minipage}
     \\
     $P\bar{4}2_1 m.1^{\prime}_c$   & $(0,0,0 | 0,0,m_z)$ $(1/2,1/2,0 | 0,0,-m_z)$ $(0,0,1/2 | 0,0,-m_z)$ $(1/2,1/2,1/2 | 0,0,m_z)$   & $(0,0,M_z)$  &
     \begin{minipage}{50mm}
          \scalebox{0.085}{\includegraphics[bb=0 0 842 617]{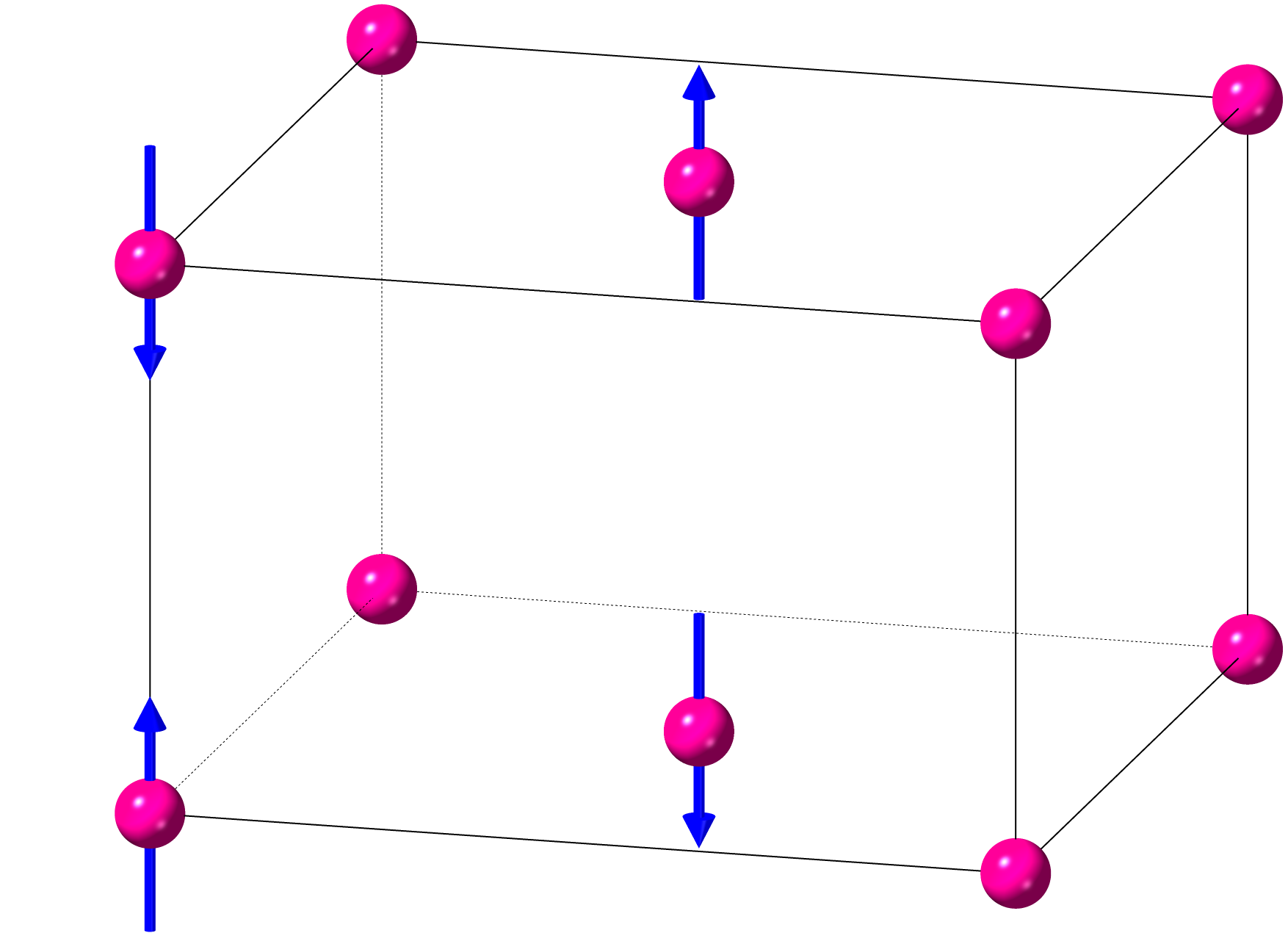}}
     \end{minipage}       
     \\          
     $Cmc2_1.1^{\prime}_c$    & $(0,0,0 | m_x,m_y,0)$ $(1/2,1/2,0 | -m_y,-m_x,0)$ $(0,0,1/2 | -m_x,-m_y,0)$ $(1/2,1/2,1/2 | m_y,m_x,0)$   & $(M_x,M_y,0)$     &
     \begin{minipage}{50mm}
          \scalebox{0.085}{\includegraphics[bb=0 0 842 617]{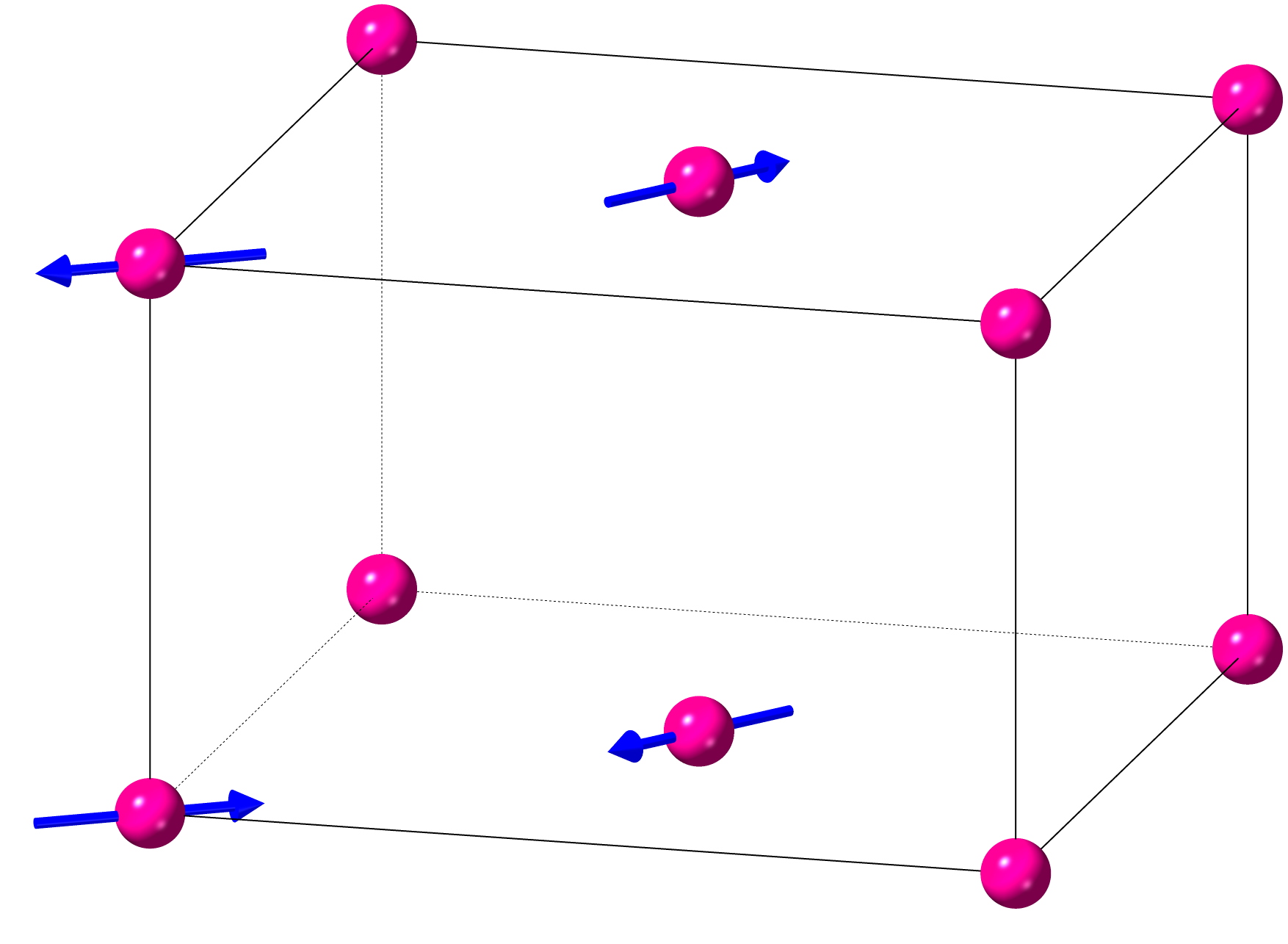}}
     \end{minipage}  
     \\
     $P2_1 2_1 2_1.1^{\prime}_c$   & $(0,0,0 | m_x,m_y,0)$ $(1/2,1/2,0 | m_x,-m_y,0)$ $(0,0,1/2 | -m_x,-m_y,0)$ $(1/2,1/2,1/2 | -m_x,m_y,0)$   & $(M_x,M_y,0)$     &
     \begin{minipage}{50mm}
          \scalebox{0.085}{\includegraphics[bb=0 0 842 617]{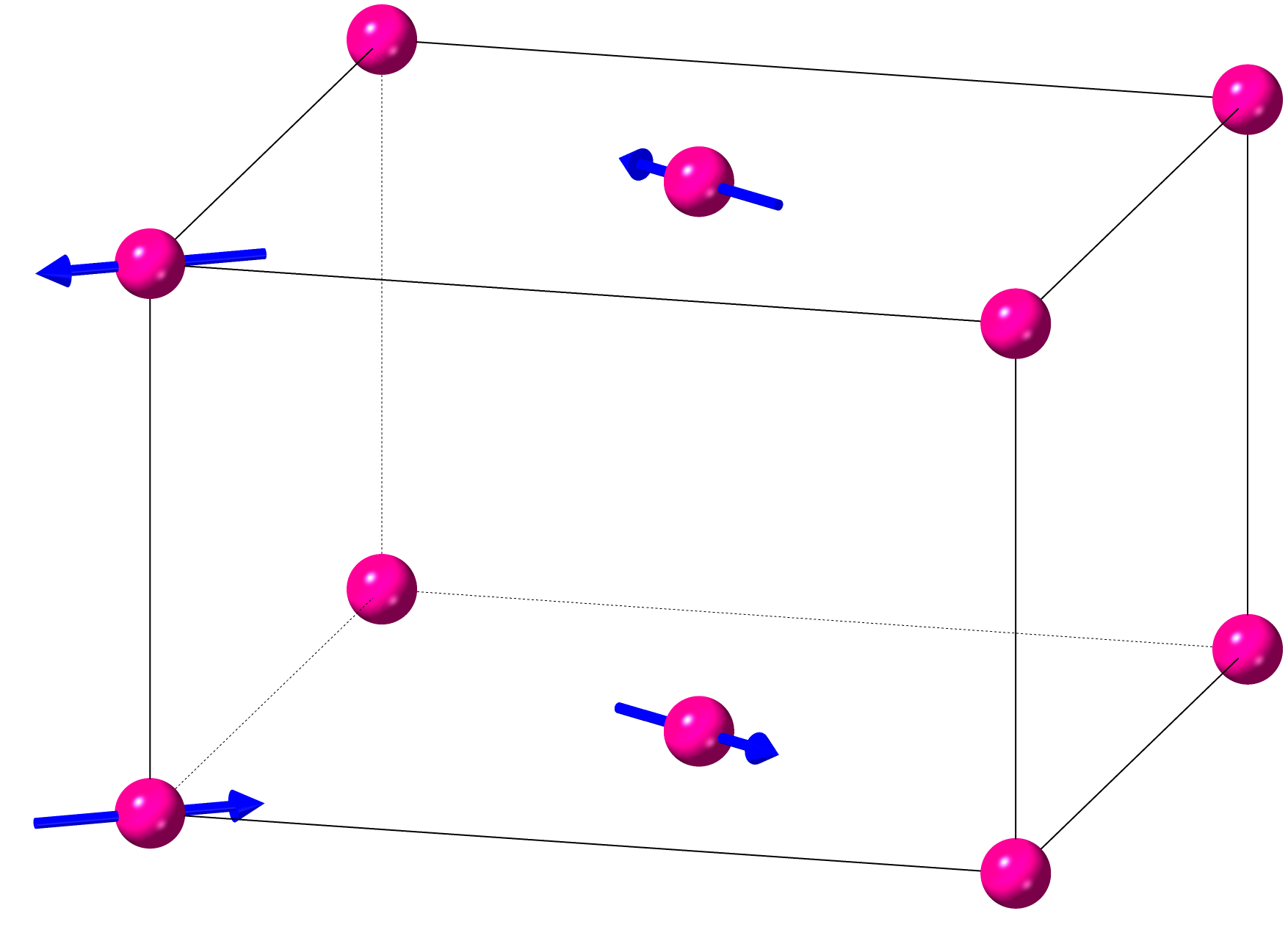}}
     \end{minipage}  
     \\
     \end{tabular}
     \label{table3}
\end{table}

\twocolumn

\onecolumn

\begin{table}
     \caption{Magnetic structure of Sr$_2$MnSi$_2$O$_7$ under magnetic space groups (MSGs) with basic information on the relations to the parent paramagnetic structure. Notations for Belov-Neronova-Smirnova (BNS) (Belov, 1957), Opechowski-Guccione (OG) (Opechowski, 1965), and Unified (UNI) symbols (Campbell, 2022) are employed, for which the BNS and OG notations differ from each other given the MSG are of type 4 having translations combined with time reversal, so-called ``anti-translations.''}
     \begin{tabular}{lcc}
     Sample  & powder     & single crystal     \\
     Temperature     & 1.70(1)~K     & 2.8(3)~K  \\
     Magnetic field & 0~T    & $\sim$0.31~T ($\parallel a$)  \\
     \hline
     Parent space group  & \multicolumn{2}{c}{$P\bar{4}2_1 m$ (\#113)}  \\
     Magnetic wavevector & \multicolumn{2}{c}{$(0,0,\frac{1}{2})$~r.l.u.}     \\
     Transformation from the parent basis    & \multicolumn{2}{c}{$(\vec{a},\vec{b},2\vec{c};0,0,0)$}     \\
     \multirow{3}{*}{MSG symbol}     & BNS: $C_c mc2_1$ (\#36.177) & BNS: $P_c 2_1 2_1 2_1$ (\#19.28)     \\
     & OG: $C_{2c}m^{\prime}m2^{\prime}$ (\#35.8.243)  & OG: $P_{2c}2_1 2_1^{\prime}2^{\prime}$ (\#18.6.118)   \\
     & UNI: $Cmc2_1.1^{\prime}_c$ (\#36.177)  & UNI: $P2_1 2_1 2_1.1^{\prime}_c$ (\#19.28)   \\
     Transformation to the standard setting & $(\vec{a}-\vec{b},\vec{a}+\vec{b},\vec{c};\frac{1}{2},0,0)$   & $(\vec{a},\vec{b},\vec{c};\frac{1}{4},0,0)$   \\
     Magnetic point group (Litvin, 2013)    & $mm2.1^{\prime}$  & $222.1^{\prime}$  \\
     \multirow{3}{*}{Unit cell parameters}     & $a=8.12170(7)$~{\AA}, $\alpha=90^{\circ}$   & $a=8.1481(67)$~{\AA}, $\alpha=90^{\circ}$   \\
     & $b=8.12177(6)$~{\AA}, $\beta=90^{\circ}$   & $b=8.1312(33)$~{\AA}, $\beta=90^{\circ}$   \\
     & $c=10.29773(6)$~{\AA}, $\gamma=90^{\circ}$   & $c=10.3198(48)$~{\AA}, $\gamma=90^{\circ}$   \\
     \multirow{4}{*}{MSG symmetry operations}  & $x,y,z,+1$ $\left\{1|0,0,0\right\}$ & $x,y,z,+1$ $\left\{1|0,0,0\right\}$ \\
     & $-x,-y,z+\frac{1}{2},+1$ $\left\{2_{001}|0,0,\frac{1}{2}\right\}$  & $x+\frac{1}{2},-y+\frac{1}{2},-z,+1$ $\left\{2_{100}|\frac{1}{2},\frac{1}{2},0\right\}$ \\
     & $y+\frac{1}{2},x+\frac{1}{2},z,+1$ $\left\{m_{1\bar{1}0}|\frac{1}{2},\frac{1}{2},0\right\}$  & $-x+\frac{1}{2},y+\frac{1}{2},-z+\frac{1}{2},+1$ $\left\{2_{010}|\frac{1}{2},\frac{1}{2},\frac{1}{2}\right\}$   \\
     & $-y+\frac{1}{2},-x+\frac{1}{2},z+\frac{1}{2},+1$ $\left\{m_{110}|\frac{1}{2},\frac{1}{2},\frac{1}{2}\right\}$     & $-x,-y,z+\frac{1}{2},+1$ $\left\{2_{001}|0,0,\frac{1}{2}\right\}$  \\
     \multirow{2}{*}{MSG symmetry centering operations}  & $x,y,z,+1$ $\left\{1|0,0,0\right\}$   & $x,y,z,+1$ $\left\{1|0,0,0\right\}$   \\
     & $x,y,z+\frac{1}{2},-1$ $\left\{1^{\prime}|0,0,\frac{1}{2}\right\}$   & $x,y,z+\frac{1}{2},-1$ $\left\{1^{\prime}|0,0,\frac{1}{2}\right\}$   \\
     Positions of magnetic atoms   & Mn1 $(0,0,0.0059(33))$   & Mn1 $(0,0,0.0000(5))$   \\
     Magnetic moments components ($\mu_{\rm B}$)  & $(M_x=2.34(34),M_y=2.15(34),0)$, $\sqrt{M_x^2+M_y^2}=3.17(34)$   & $(M_x=0.31(93),M_y=3.14(2),0)$, $\sqrt{M_x^2+M_y^2}=3.15(9)$   \\
     \multirow{10}{*}{Positions of nonmagnetic atoms}  & Sr1\_1 $(0.8284(8),0.3284(8),0.2507(38))$    & Sr1 $(0.8342(3),0.3334(2),0.2469(2))$   \\
     & Sr1\_2 $(0.3400(8),0.1600(8),0.7557(38))$    & Si1 $(0.6370(5),0.1375(3),0.0307(2))$   \\
     & Si1\_1 $(0.6385(18),0.1385(18),0.0407(44))$  & O1 $(0,1/2,0.4161(3))$     \\
     & Si1\_2 $(0.1370(18),0.3630(18),0.9768(45))$  & O2 $(0.6404(4),0.1396(2),0.3752(2))$     \\
     & O1\_1 $(0,1/2,0.4135(38))$                    & O3\_1 $(0.0779(3),0.1958(2),0.3954(2))$   \\
     & O1\_2 $(1/2,0,0.5805(37))$                    & O3\_2 $(0.1968(3),0.9218(2),0.6045(2))$   \\
     & O2\_1 $(0.6447(13),0.1447(13),0.3765(42))$    &    \\
     & O2\_2 $(0.1322(11),0.3679(11),0.6285(41))$    &    \\
     & O3\_1 $(0.0842(9),0.1991(13),0.4014(38))$     &    \\
     & O3\_2 $(0.1901(13),0.9288(10),0.6089(37))$    &    \\
     \end{tabular}
     \label{table4}
\end{table}

\twocolumn

\end{document}